\documentclass{elsart}

\begin{document}

 \renewcommand              
 \baselinestretch 1         
 \baselineskip 16pt         

\begin{frontmatter} 

\title{Determination of the order of phase transitions in Potts model
       by the graph-weight approach}  

\author{Zvonko Glumac\thanksref{corresp}}, 
\thanks[corresp]{Corresponding author. Tel.: (385)-1-4608-211;
                                       fax:  (385)-1-4680-399; 
                                       e-mail: zvonko@athena.ifs.hr. } 
\author{Katarina Uzelac}

\address{ Institute of Physics, P.O.B. 304,
Bijeni\v{c}ka 46, HR-10000 Zagreb, Croatia}

\begin{abstract}
 We examine the order of the phase transition in the Potts model by using 
the graph representation for the partition function, which allows treating a 
non-integer number of Potts states. 
The order of transition is determined by the analysis of the shape of the 
graph-weight probability distribution.
The approach is illustrated on special cases of the one-dimensional Potts 
model with long-range interactions and on its mean-field limit. 
\\

\end{abstract}


\end{frontmatter}

\section{Introduction}

One of the less trivial questions in the study of phase-transitions is the 
determination of the order of transition. 
A large amount of work, especially on the Potts \cite{P52} model, ranging from 
the exact solutions \cite{KMS54,B73} to  different kinds of Monte Carlo (MC) 
simulations \cite{B81,S83,SW87,FS88,LK91a,LK91b,JK95,LB95,JV97} (to cite 
only some of them), has been done in order to establish reliable and at the 
same time applicable criteria for distinction between the first- and second-order 
phase transition.

The coexistence of two different phases at the first-order phase transition point 
and its absence at the second-order phase transition point is the main physical 
fact on which all of the above methods (including the present one) are based. 
Performing the MC simulations on finite models in order to obtain the probability 
distribution of some quantity (such as the energy or the order parameter) which has 
different values in different phases, 
is the usual way to investigate  these phase transitions numerically\cite{BH92}.  
In the present paper we show that the quantity which we define as the graph 
weight also has different values in the coexisting phases at the first-order 
transition point.  
In Section \ref{sec-res} it is shown how the graph weight is related to the 
energy and the free energy of the system. 
The coexistence of phases with different energy at the first-order 
transition point then implies the coexistence of graphs with different weights. 
It can be identified by the two peaks in the graph-weight probability distribution.        

In order to test the graph approach, we analyse the Potts model in two cases: 
the mean-field (MF) case defined by the interactions of equal strength between all 
particles of the model and the one-dimensional $(1d)$ model with long-range 
(LR) interactions decaying with distance $r$ as $r^{-(1+\sigma)}$, $\sigma>0$. 
In both cases two regimes (of the first- and second-order transition) are present. 
The regimes are separated by a point $q_c$ in the MF case and by a line 
$(q_c, \sigma_c)$ in the LR case.  

The present approach has also the advantage of dealing directly with the 
non-integer values of the Potts states $q$. (Note that in reference \cite{LK91a} 
the non-integer $q$ values were not obtained by a direct calculation, but by  a 
histogram-like extrapolation from the results for integer-$q$.) 
This may be of interest in studying the threshold $q_c$, separating the first- from 
the second-order transition regime, which is not always an integer. 
For example, in three-dimensional short-range (SR) Potts model, $q_c$ is  not 
an integer \cite{LK91a,BS79}, and in $(1d)$ Potts model with LR interaction, 
 the onset of the first order transition at $q_c$ seems to depend continuously on the
interaction-range parameter $\sigma$ \cite{UG97,GU98}.

The plan of the paper is as follows. 
In the next section we define the model and introduce the corresponding graph 
expansion. 
The basic steps of MC algorithm are also explained in that section. 
In the third section we discuss separately the results for the MF and the LR 
interaction case. 
In the last section we summarize and discuss the advantages and open problems 
connected with the graph approach.  


\section{Model and method} 
\label{sec-mm} 

The graph representation \cite{FK69} was already used in MC investigation of
critical behaviour of Potts models \cite{S83,SW87,JK95}.   
In the present approach we follow the line of reasoning given in reference 
\cite{S83}, but with the basic difference that we study directly the probability 
distribution of graph weights and not the cluster probability distribution. 

We begin by rewriting the model in the graph language. 
The reduced Hamiltonian of the model, with periodic boundary conditions,
has the form 
\begin{equation}
\frac{-\, H}{k_B\, T} \; = \; \sum_{i = 1}^{N-1} \; \sum_{j = 1}^{N-i} \;
 K_j \; \delta (s_i, s_{i+j}).  \label{eq:ham} 
\end{equation}
 where $s_i = 0, 1, \dots, q-1$ is the Potts particle placed on the $i$-th site 
of the chain, while $\delta$ denotes the Kronecker symbol. 
$K_j$'s denote interactions between two particles at distance $j$. 
Due to boundary conditions in the LR case, each $K_j$ has two contributions 
involving interactions at distances $j$ and $(N-j)$. 
In the MF case, the interactions are unique and the boundary conditions have no 
meaning. 
We use the system of units where $K_j$ is proportional to the inverse temperature. 

The usual substitution,  
$\exp[ K_j \, \delta(s_i, s_{i+j}) ] = 1 + v_j \, \delta(s_i, s_{i+j})$, 
with $v_j = \exp(K_j) - 1$, leads to the partition function of the form 
\begin{equation} \label{eq:pf} 
Z_N =\prod_{l = 1}^N \;\sum_{s_l = 0}^{q-1}\;\;\exp (-H/ k_B T) = 
     \prod_{l = 1}^N \;\sum_{s_l = 0}^{q-1}                               
\;\; \prod_{i = 1}^{N-1}\prod_{j = 1}^{N-i}\; [1+v_j\delta(s_i, s_{i+j})].
\end{equation}
It is straightforward to establish one-to-one correspondence between each 
member of the r.h.s. of the above equation and a graph on the chain consisting 
of $N$ particles. 
Each square bracket can contribute  to the above product  in two ways:
it gives $1$ when there is no connection between the $i$-th and the $i+j$-th particle 
(inactive link), and $v_j\,\delta(s_i, s_{i+j})$ when there is a direct connection 
(active link) between these two particles. 
The analytical expression attached to each graph ${\mathcal G}$ will be called the graph 
weight, $W_N({\mathcal G})$. 
The explicit expression for $W_N({\mathcal G})$ follows from the r.h.s. of eq. (\ref{eq:pf}).  
We can describe every graph consisting of $c({\mathcal G})$ clusters.
By {\it cluster} we mean a set of particles interconnected by any type of active 
links and disconnected from other particles. 
Single particles are considered as one-particle clusters.
The products of $\delta$-functions from the expression for each graph will delete 
all except $c$ $q$-summations on the r.h.s. of (\ref{eq:pf}), so that the analytical 
expression for $W_N({\mathcal G})$ is 
\begin{equation}
\label{eq:W} 
 W_N ({\mathcal G}) = v_1^{b_1({\mathcal G})} \; v_2^{b_2({\mathcal G})} \; 
\dots v_{N-1}^{b_{N-1}({\mathcal G})} \;
q^{c({\mathcal G})}. 
\end{equation}
The symbols $b_j({\mathcal G})$ denote the total number of active links of type $j$. 

In this way, the summation over all $2^{N(N-1)/2}$ possible graphs between $N$ 
particles corresponds to the partition function of the model  
\begin{equation}
Z_N = \sum_{{\rm all} \; {\mathcal G}} W_N({\mathcal G}) = 
      \sum_{ \; W} {\mathcal N}_{N, W} \; W_N.  \label{eq:pfallW} 
\end{equation}
The ${\mathcal N}_{N, W}$ denotes the number of different graphs with the same weight 
$W_N$. 
The above equation is thus the graph analogy of the more familiar expression 
\begin{equation} \label{eq:pfE} 
Z_N = \sum_{\; E} {\mathcal N}_{N, E} \; e^{-E_N/T}, 
\end{equation}  
where the summation runs over all different energies $E_N$ of the system, 
while ${\mathcal N}_{N, E}$ is their degeneracy. 
Indeed, behind the analogy, there is a connection between the average number 
of clusters and the active links and average of energy and free energy of the system. 
The derivative of the partition function, given by eqs. (\ref{eq:W}), 
(\ref{eq:pfallW}) and (\ref{eq:pfE}), over temperature \cite{JK95} leads to        
\begin{equation} \label{eq:poT}
\;\left\langle\;E_N\;\right\rangle = T^2\,\sum_{j=1}^{N-1}
\frac{v_j+1}{v_j}\;\left\langle\;b_j\;\right\rangle
\,\frac{\partial K_j}{\partial T},
\end{equation}  
while the derivative over $q$ leads to 
\begin{equation} \label{eq:poq}
\left\langle\;\frac{\partial \ln {\mathcal N}_{N, E}(q)}{\partial q}\;
\right\rangle - 
\frac{1}{T}\;\left\langle\;\frac{\partial E_N}{\partial q}\;\right\rangle 
\,=\, \frac{1}{q}\;\langle\;c\;\rangle.  
\end{equation}  
 The coexistence of two phases characterized by different values of energy  
corresponds, by the above relations, to the coexistence of graphs with different 
weights. 

According to (\ref{eq:pfallW}), one introduces the graph-weight probability 
distribution $P_N$  
\begin{equation}
P_N = \frac{{\mathcal N}_{N, W} \; W_N}{Z_N}. 
\end{equation}
Numerically, $P_N$ is obtained by a simple MC simulation of Metropolis type. 
The basic steps are:   

(a) Pick at random one link in the graph ${\mathcal G}$ and change its status 
    from active to inactive or {\it vice versa}. The resulting graph is 
    called ${\mathcal G}'$.  \\
(b) Compare the random number $0 < r \leq 1$ with the ratio 
    $W_N({\mathcal G}')/W_N({\mathcal G})$. 
    If the ratio is smaller than $r$ keep ${\mathcal G}$, otherwise save ${\mathcal G}'$
    as ${\mathcal G}$. \\
(c) Return to (a).

    At each step, the number of clusters has to be counted.
Since we deal with a model with interactions of infinite range, the determination 
of clusters could not be performed by determination of their boundaries, but one 
has to check for possible connections with all the particles of the considered cluster.
To do so, during the simulations, we have to keep record of the cluster structure 
of the system, i.e. which particle belongs to which cluster. 
We start with a simple and known cluster configuration. 
In each of the following MC steps, when one link is changed, we examine whether 
this has produced a change in cluster structure:
if the link is deactivated - whether the corresponding cluster is split in two or not;
if the link is activated - whether two clusters get connected or not.

Counting graphs by their weights through a large number of steps gives the 
unnormalized graph-weight probability distribution. 
Unfortunately, the shortcoming of the graph approach, is that it takes a much 
longer time to obtain the same precision of results compared to the simulation 
techniques that can be applied only for integer $q$ (like algorithm by Luijten 
and Bl\"ote \cite{LB95}). 
 On the other hand, the comparison of the present approach with the simple 
Metropolis single spin flip algorithm done for the model of N particles with 
integer $q$  shows that $10^4$ flips per link of everyone of $N(N-1)/2$ links in 
graph simulations give approximately the same precision as $10^6$ one-particle 
flips per particle in the spin simulations.
For the chain of 400 sites considered here, the graph approach requires about 
$2-5$ times larger CPU time than the Metropolis algorithm on spins. 

\section{Results} \label{sec-res}   

The simulations were applied to the two mentioned cases of the Potts model with 
$N = 100, 150, 200, 250, 300, 350$ and $400$ particles. 
 In final extrapolations only the data for $N\geq 200$ were used.
The precision of simulations is determined by performing $10^4$ flips per link. 

\subsection{Mean-field case} \label{subsec-mfres} 

The MF case of Hamiltonian (\ref{eq:ham}) is given by  taking the equal 
strength of interactions among all the particles, i.e. $K_j = K\, / \, N$, where 
$K$ denotes the inverse temperature. 
The exact work of Kihara {\it et al.} \cite{KMS54} puts in evidence that the MF 
Potts model has a  second-order phase transition for $q\leq q_c=2$ and a
first-order phase transition for $q > 2$. 
It thus qualifies as a good example for testing the present approach. 
In our simulation we start with the case $q = 3$. In continuation, we consider 
lower values of $q$, closer to the threshold $q_c$ in order to examine how 
efficient our approach can be in detecting weak first-order transitions.  

The shape of the finite-$N$ graph-weight probability distribution $P_N$ depends 
on temperature in a similar way as the energy-probability distribution does.  
Far from the transition temperature, $P_N$ has a gaussian form. 
At the transition, the shape of $P_N$ depends on the order of the transition.
For second-order transitions, the shape has a non-gaussian form, but it still 
has only one maximum. 
For first-order transitions, $P_N$ has two maxima which transform into two 
$\delta$-functions in the thermodynamic limit. 
We define the temperatures $T_N$ where the two maxima in $P_N$ are of equal 
height. 
The corresponding positions on the graph-weight axis are $W_N^{\rm o}$ (for the
ordered phase) and $W_N^{\rm do}$ (for the disordered phase).  
When the two-peak structure of $P_N$ becomes more pronounced with increasing 
$N$, one may conclude that the transition is of the first order in the thermodynamic 
limit with two coexisting phases: 
the one described by the graph weight  $W_N^{\rm o}\to W^{\rm o}$, stable below 
the transition temperature $T_N\to T_t$, and the other one described by 
$W_N^{\rm do}\to W^{\rm do}$ stable above $T_t$.

In Figure 1 are presented the results of simulations for $q=3$. 
It shows dependence of $P_N$ on $w_N\,\equiv\,\ln\,W_N\,/\;N\,\ln\,N$ at 
temperatures $T_N$.  
The peaks emerge for $200 \leq N \leq 400$ and become more pronounced with 
increasing $N$.
By the above arguments, the $N$-dependence of $P_N$ confirms the existence of 
the first-order phase transition, in agreement with the exact solution \cite{KMS54}. 

The behaviour for smaller $N$ considered ($N<200$) points out that only one 
maximum in $P_N$ may have different origins: 
either the transition is of the second order in the thermodynamic limit and we 
observe its finite-$N$ behaviour, or the transition is of the first order in the 
thermodynamic limit, but the correlation length, although finite, is comparable 
with the system size used and we are not able to see the coexistence of the two 
phases. 
Arbitrarily close to the threshold which separates the first- from the 
second-order transitions, the correlation length is expected to become 
arbitrarily large and finally whatever system size is used, it will always be too 
small to show the first-order character of the transition. 
Such a situation is expected in the MF case when $q$ approaches 2. 
The simulations on $q = 2.8$ model performed  at the corresponding 
temperatures defined as $T_N$  give  the distributions similar to those shown 
in Fig. 1. 
On the contrary, simulations performed for $q=2.1$  and $2.5$ do not show the 
two-peak structure in the graph-weight probability distribution for the considered 
system sizes.             

To some extent, the above presented simulations can be compared to the exact 
solution. 
Since ${\mathcal N}_{N, E}(q)$ and $E_N(q)$ can be exactly calculated for the 
transition temperature $T_{\rm MF}^{-1}\,=\,2\,[(q - 1)/(q - 2)] \ln\,(q - 1)$,
\cite{KMS54}, it is easy to obtain from eqs. (\ref{eq:poT}) and (\ref{eq:poq}) 
the analytic expressions for the positions $\langle\,w_N^{\rm o, do}\,\rangle$ 
of the two peaks at temperature $T_{\rm MF}$ in the large $N$ limit: 

\begin{eqnarray} \label{eq:wodo} 
\left\langle\,w_N^{\rm o}\,\right\rangle & = &
 -\; \frac{(q-1)^3 + 1}{q^2\;(q-2)}\;\ln (q-1)  
 +  \frac{a_q^{\rm o}}{\ln N}\;  + {\mathcal O}(\frac{1}{N}),\nonumber \\
\langle\,w_N^{\rm do}\,\rangle & = &
 -\; \frac{q-1}{q\;(q-2)}\;\ln (q-1)
 +  \frac{a_q^{\rm do}}{\ln N}\;  + {\mathcal O}(\frac{1}{N})\, ,
\end{eqnarray}
where \\
\lefteqn{a_q^{\rm o} = 
\frac{(q-1)^3 + 1}{q^2\;(q-2)}\,\ln
(q-1)\,\ln\left[2\frac{q-1}{q-2}\,\ln(q-1)\right]
+ \frac{\ln q}{q-1} \left[ 1 - \frac{\ln (q-1)}{q(q-2)} \right], 
}       \\ 
\lefteqn{a_q^{\rm do} = 
\frac{q-1}{q\;(q-2)}\,\ln (q-1)\,\ln\left[2\frac{q-1}{q-2}\,\ln(q-1)\right]
+ \ln q \left[ 1 - \frac{(q-1)\ln (q-1)}{q(q-2)} \right]\, .
} 

In Table 1 we compare the data for $q=3$ obtained by MC simulations at 
$T_{N=400} = 0.3566$, with those calculated from (\ref{eq:wodo}) for finite 
$N=400$ and for $N \to\infty$.  

\begin{table}[hbt]   
\caption{Simulated (MC) and exact values of the 
positions $\left\langle\,w_N^{\rm o}\,\right\rangle$ and
$\langle\, w_N^{\rm do}\,\rangle$ of two maxima in graph-weight
probability distribution of $q=3$ MF Potts model.}
\begin{center} 
\begin{tabular}{ ccccc } 
\hline 
T & N & Method & $\left\langle\,w_N^{\rm o}\,\right\rangle$ &
                 $\langle\,w_N^{\rm do}\,\rangle$ \\  
\hline 
 $T_N$ & $400$ & MC & -0.5271 & -0.3123 \\
 $T_{\rm MF}$ & $400$ & exact & -0.5045 & -0.2847 \\
 $T_{\rm MF}$ & $\infty$ & exact & -0.6931 & -0.4621 \\
\hline
\end{tabular}  
\end{center} 
\end{table} 

The values of $\langle\,w_N^{\rm o, do}\,\rangle$ cited in Table 1 show that the 
simulated results fit well the exact values for finite $N = 400$. 
Relatively high difference ($24\%$ for $\langle\,w_N^{\rm o}\,\rangle$ and 
$32\%$ for $\langle\,w_N^{\rm do}\,\rangle$) compared to the exact $N\to\infty$ 
values comes from the slow $1/\ln\,N$ convergence. 

\begin{table}[hbt]
\caption{Values of MF temperatures $T^{\rm MF}_N$ and LR temperatures 
$T^{\rm LR}_N$ at which $P_N$ shows two peaks of approximately same heights, 
followed by the extrapolated values. }   
\begin{center}
\begin{tabular}{ clll }
\hline
 N & $T^{\rm MF}_{N, q = 2.8}$ & $T^{\rm MF}_{N, q = 3}$& 
     $T^{\rm LR}_{N, q = 3, \sigma = 0.1}$ \\  
\hline
  $200$ & $0.3694$ & $0.3533$ & $3.25248$ \\
  $250$ & $0.37042$& $0.35491$& $3.3402$ \\
  $300$ & $0.3699$ & $0.35490$& $3.4235$ \\
  $350$ & $0.37298$& $0.35615$& $3.4880$ \\
  $400$ & $0.3730$ & $0.3566$ & $3.55304$ \\
\hline
  $\infty$ & $0.376$ & $0.360$ & $7.73$ \\
\hline
\end{tabular}
\end{center}
\end{table}

In Table 2 we present the corresponding data for $T_N$ (obtained in 
simulations) in function of size. 
Since for large $N$ the dependence of $T_N$'s is approximately linear in $N^{-1}$,
we use the extrapolation form $T_N = T_{{\rm MF},t} + b/N$. 
The coefficients $T_{{\rm MF},t}$ and $b$ were obtained in least-squares
approximation (LSA).  
 By taking into account only the data for sizes $200 \leq N \leq 400$, one obtains 
$T_{{\rm MF},t}(q=2.8) = 0.376$ and $T_{{\rm MF},t}(q=3.0) = 0.360$. 
In both cases, the difference between the extrapolated values and the exact ones 
$T_{\rm MF}(q=2.8)\simeq 0.378$ and $T_{\rm MF}(q=3.0) \simeq 0.361$, 
is less than $1\%$. 


\subsection{One-dimensional long-range case}

In the case of one-dimensional Potts model with interactions decaying as 
$1/r^{1+\sigma}$, the interaction strength $K_j$ is given by                  
$K \, [1\, / \,j^{1+\sigma} + 1\, / \,( N - j)^{1+\sigma}]$, where $K$ denotes 
the inverse temperature. 
This model has a phase transition at finite temperature \cite{ACCN88,GU93} for 
all $q$ and $0<\sigma\leq 1$. 
Our recent MC simulations of the energy probability distribution on $q = 3$ and 
$q = 5$ models \cite{UG97,GU98} have shown that the order of the transition 
depends on $q$ and $\sigma$. 
For the same fixed value $q=3$, it was shown that the transition changes from the
first- to the second-order one with increasing $\sigma$.         

Again, according to the relations (\ref{eq:poT}) and (\ref{eq:poq}), one can 
interpret the meaning of the graphs through their connection with the energy and 
the free energy of the system. 

The purpose of simulations presented in this subsection is twofold:
first, we wish to test the above described graph approach comparing the results 
with those obtained earlier by MC simulation of energy probability distribution 
(strong first-order transition for model with $q = 3$ and $\sigma = 0.1$);
second, by using the same graph approach we wish to analyze an example 
of a non-integer $q$ model with a second-order phase transition \cite{A78,AP79} 
(we choose $q = 0.5$ and $\sigma= 0.1$ and $0.8$). 

The result of simulations for the case with $q = 3, \sigma = 0.1$ is shown in Fig. 2. 
It presents $P_N$ versus $w_N$ for system sizes ranging from $200$ to $400$,   
at temperatures $T_N$ where two peaks of the approximately same height 
appear. 
The figure shows that the depth of the minima increases with $N$.   
According to the discussion in the preceding subsection, a behaviour like that 
is characteristic for a first-order phase transition. 
This conclusion is confirmed by the results\cite{GU98} of the MC simulations for 
the energy probability distribution. 

 The temperatures $T_N$, where $200 \leq N \leq 400$ are presented in Table 2. 
Compared to the MF results and also to those for higher values of $\sigma$, the results for 
$\sigma = 0.1$ converge very slowly, so that, for sizes considered here, we do not 
expect so good accuracy in extrapolation to $N\to\infty$.
The extrapolation which takes one correction term, of the form $T_N = T_{LR, t} + b/N^x$ 
gives $T_{LR, t} = 7.73$ with the convergence exponent $x=0.1$, while taking the form with 
two correction terms, e.g.  $T_N = T_{LR, t} + b/N^x + c/N^{2 \cdot x}$ gives by LSA fit  $x = 0.12$ 
and  $T_{LR, t} = 7.45$, value which differs by $4\%$.
The first result for $T_{LR, t}$ shows the discrepancy of $8\%$ compared to the 
improved finite-range scaling (FRS) \cite{GU93} result $(T^{\rm FRS} = 7.14)$ 
cited in  \cite{GU98}, and even larger  discrepancies when compared to the 
renormalisation-group (RG) result $(T^{\rm RG} = 6.72)$ \cite{CM97} and the 
earlier MC result  $(T^{\rm MC} = 6.25)$ \cite{GU98}.  
(Notice however that, due to the same slow convergence, the difference between 
these previous results is also quite significant when $\sigma = 0.1$.)
Apart from the reduced precision due to the small convergence exponent, 
the additionnal source of deviation can come from the crossover effect and be the 
consequence of the limitation to relatively small sizes for which the system does not 
exhibit full qualities of the first-order transition. 

In order to investigate the case with non-integer $q$, we also considered 
the case $q = 0.5$ with $\sigma = 0.1$ and $0.8$, where by \cite{A78} and 
\cite{AP79} the transitions are of classical and non-classical second-order 
type, respectively. 
The simulations were performed for a wide range of temperatures around critical 
temperatures obtained earlier by FRS \cite{GU93}.                   
As one can expect on the grounds of the discussion at the beginning of the
preceding subsection, no double-peak structure in graph-weight probability 
distribution has been found. 


\section{Conclusion}

The graph-weight probability distribution is introduced and applied to the 
analysis of the order of transition on two special cases of the Potts model:  
its mean-field case and the case in $1d$ with power-law decaying interactions. 
The mean-field case was analyzed for $q = 3$ and several non-integer
values approaching $q_c = 2$. The long-range case was examined for $q = 3$ 
and $q = 0.5$ with $\sigma = 0.1$ and $\sigma = 0.1, 0.8$, respectively.  
The simulations were limited by time to the systems of sizes $N \leq 400$.      

It is shown that the graph weight is an appropriate quantity to distinguish the 
coexisting phases at the first-order transition point. 
The physical interpretation of the graph weight becomes transparent through the 
equations (\ref{eq:poT}) and (\ref{eq:poq}) which relate the average number of 
active links and clusters to the average of energy and the free energy of the system.

By analyzing the graph-weight probability distributions of the MF and $1d$ LR 
Potts models, we have observed the behaviour of distributions characteristic for 
first-order transitions in the case of $q = 3$ and $2.8$ MF models and $q=3, 
\sigma = 0.1$ LR model, while the transition of the second order was obtained in 
the case of $q=0.5, \sigma = 0.1, 0.8$ LR model.

Transition temperatures have also been  calculated. The estimated 
values of those temperatures in the thermodynamic limit agree for the MF case 
values within $1\%$, while for the case with power-law decaying interactions, 
where only the approximate results are available, it agrees with the discrepancy
of $5\%$ to $24\%$ depending on the method. 

All of the above facts  qualify the graph-weight probability distribution as
an alternative quantity in investigations of the order of transitions in Potts models,
capable to deal directly with non-integer values of $q$.   
This makes the graph approach interesting in the framework of efforts of 
determination of the border between the first- and second-order regions by 
continuously varying $q$ in the $(q, d)$ plane of SR models or $(q, \sigma)$ 
plane in $1d$ LR models.  



  \newpage
 \vspace{2cm} 

\centerline{\Large Figure captions}  

\vspace{2cm}  

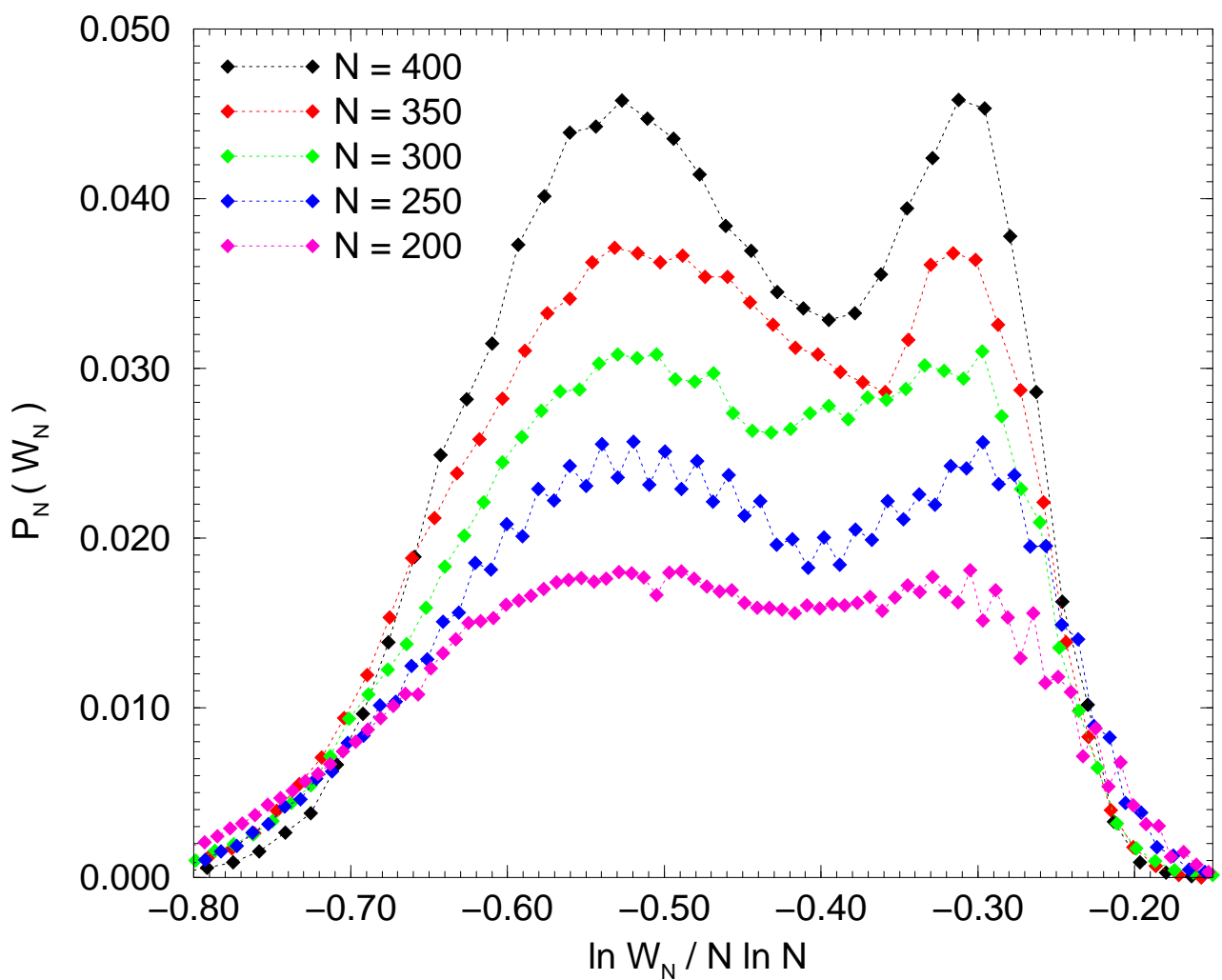
{\bf Fig. 1}: The simulation of the graph-weight probability distribution $P_N$ 
of the $q=3$ MF Potts model versus $w_N \equiv \ln\,W_N\;/\;N\;\ln\,N$ performed 
at temperatures $T_N$ (see in text).  

\vspace{0.5cm} 

{\bf Fig. 2}: The simulation of graph-weight probability distribution $P_N$ of the 
$q=3, \sigma = 0.1$ $1d$ LR Potts model versus $w_N \equiv \ln\,W_N\;/\;N\;\ln\,N$ 
performed at temperatures $T_N$ (see in text).


\newpage 

\begin{thebibliography}{9} 

\bibitem{P52} R.B. Potts, Proc. Camb. Phil. Soc. {\bf 48} (1952) 106. 

\bibitem{KMS54} T. Kihara, Y. Midzuno and T. Shizume,
              J. Phys. Soc. Jpn. {\bf 9} (1954) 681. 

\bibitem{B73} R.J. Baxter, J. Phys. C {\bf 6} (1973) L445.

\bibitem{B81} K. Binder, Z. Phys. B {\bf 43} (1981) 119.

\bibitem{S83} M. Sweeny, Phys. Rev. B {\bf 27} (1983) 4445. 

\bibitem{SW87} R.H. Swendsen and J.-S. Wang, Phys. Rev. Lett. {\bf 58} (1987) 86.

\bibitem{FS88} A.M. Ferrenberg and R.H. Swendsen, Phys. Rev. Lett. {\bf 61} (1988) 2635.

\bibitem{LK91a} J. Lee and J.M. Kosterlitz,  Phys. Rev. B {\bf 43} (1991) 1268 .

\bibitem{LK91b} J. Lee and J.M. Kosterlitz,  Phys. Rev. B {\bf 43} (1991) 3265 .

\bibitem{JK95} W. Janke and S. Kappler, Phys. Rev. Lett. {\bf 74} (1995) 212.

\bibitem{LB95} E. Luijten and H.W.J. Bl\"ote, Int. J. Mod. Phys. C {\bf 6} (1995) 359.

\bibitem{JV97} W. Janke and R. Villanova, Nucl. Phys. B {\bf 489} [FS] (1997) 679. 

\bibitem{BH92}  K. Binder and H.J. Herrmann, in {\it Monte Carlo Simulation in
                Statistical Physics}, eds. M. Cardona, P. Fulde, K. von Klitzing, 
                and H.-J. Queisser (Springer-Verlag, Berlin, 1992). 

\bibitem{BS79} see, e.g., H.W.J. Bl\"ote and R.H. Swendsen, Phys. Rev. Lett.
               {\bf 43} (1979) 799 and references in: 
               F.Y. Wu,  Rev. Mod. Phys. {\bf 54} (1982) 235.  

\bibitem{UG97} K. Uzelac and Z. Glumac, Fizika B {\bf 6} (1997) 133.

\bibitem{GU98} Z. Glumac and K. Uzelac, Phys. Rev. E {\bf 58} (1998) 4372 . 

\bibitem{FK69} P. W. Kasteleyn and C. M. Fortuin, J. Phys. Soc. Jpn. 
               Suppl. {\bf 26} (1969) 11; 
               C. M. Fortuin and P. W. Kasteleyn, Physica {\bf 57} (1972) 536.   

\bibitem{ACCN88} M.Aizenman, J.T. Chayes, L. Chayes and C.M. Newman, 
                 J. Stat. Phys. {\bf 50} (1988) 1.  

\bibitem{GU93} Z. Glumac and K. Uzelac, J. Phys. A {\bf 26} (1993) 5267. 

\bibitem{A78} A. Aharony, J. Phys. C {\bf 11} (1978) L457.

\bibitem{AP79} A. Aharony and P. Pfeuty, J. Phys. C {\bf 12} (1978) L125.

\bibitem{CM97}  S.A. Cannas  and A.C.N. de Magalh\~aes, J. Phys. A {\bf 30} (1997) 3345. 

\end{thebibliography}
\end{document}